\documentclass[twocolumn,superscriptaddress,nofootinbib]{revtex4-1}
\usepackage{multirow}
\usepackage{array}
\usepackage{xcolor}
\newcolumntype{C}[1]{>{\centering\let\newline\\\arraybackslash\hspace{0pt}}m{#1}}
\usepackage{comment}
\usepackage{slashed}
\usepackage{latexsym}
\usepackage{graphics}
\usepackage{bm}
\usepackage{color}
\usepackage{amsmath}
\usepackage{amssymb}
\usepackage{feynmp-auto}
\usepackage{slashed}
\makeatletter
\newcommand\footnoteref[1]{\protected@xdef\@thefnmark{\ref{#1}}\@footnotemark}
\makeatother
\usepackage{color,hyperref}
\hypersetup{colorlinks,breaklinks,
 linkcolor=blue,urlcolor=blue,
 anchorcolor=blue,citecolor=blue}

\usepackage{tikz}
\usetikzlibrary{shapes.geometric,decorations.fractals,shadows,shapes.multipart,arrows,snakes,positioning,arrows}
\usetikzlibrary{decorations.pathmorphing,decorations.markings}
\usetikzlibrary{calc,intersections}
\begin{document}
 
\title{Metastable Nuclear Isomers as Dark Matter Accelerators}
\author{Maxim Pospelov}
\affiliation{Perimeter Institute for Theoretical Physics, Waterloo, ON N2J 2W9, Canada}
\affiliation{Department of Physics and Astronomy,
University of Victoria, Victoria, BC V8P 5C2, Canada}
\author{Surjeet Rajendran}
\affiliation{Department of Physics \& Astronomy, The Johns Hopkins University, Baltimore, MD 21218, USA}
\author{Harikrishnan Ramani}\email{hramani@berkeley.edu}
\affiliation{Berkeley Center for Theoretical Physics, Department of Physics,
University of California, Berkeley, CA 94720, USA}
\affiliation{Theoretical Physics Group, Lawrence Berkeley National Laboratory, Berkeley, CA 94720}
\begin{abstract}
Inelastic dark matter and strongly interacting dark matter are poorly constrained by direct detection experiments since they both require the scattering event to deliver energy from the nucleus into the dark matter in order to have observable effects. We propose to test these scenarios by searching for the collisional de-excitation of meta-stable nuclear isomers by the dark matter particles. The longevity of these isomers is related to a strong suppression of $\gamma$- and $\beta$-transitions, typically inhibited by a large difference in the angular momentum for the nuclear transition. The collisional de-excitation by dark matter is possible since heavy dark matter particles can have a momentum exchange with the nucleus comparable to the inverse nuclear size, hence lifting tremendous angular momentum suppression of the nuclear transition. This de-excitation can be observed either by searching for the direct effects of the decaying isomer, or through the re-scattering or decay of excited dark matter states in a nearby conventional dark matter detector setup. Existing nuclear isomer sources such as naturally occurring $^{180m}$Ta, $^{137m}$Ba produced in decaying Cesium in nuclear waste, $^{177m}$Lu from medical waste, and $^{178m}$Hf from the Department of Energy storage can be combined with current dark matter detector technology to search for this class of dark matter. 
\end{abstract}

\maketitle

\section{Introduction}

The nature of dark matter (DM) is currently one of the pivotal issues in particle physics and fundamental physics in general. Particles with a mass around the weak scale are natural candidates for DM since these arise in a large class of theories that attempt to solve other problems of the Standard Model (SM). Direct detection (DD) experiments that probe the elastic scattering of weak scale DM with nuclei have set stringent constraints on the existence of such particles, and significantly limit scenarios where DM is composed of weakly interacting massive particles (WIMPs). Most of the DD experiments targeting these types of models 
require the DM to transfer a fraction of its kinetic energy to nuclei/electrons of the detector material. 
Thus, these experiments can only probe DM particles that are sufficiently heavy and fast so that they have enough kinetic energy to dump into a detector. (Besides elastic scattering, these experiments can also probe metastable excited DM states that can de-excite in collisions with nuclei and deposit DM excitation energy into the detector.)

Constraints on DM scattering are dramatically weakened in several special cases.
The first exception is the case of strongly-interacting particles. If DM-nucleus scattering cross section is relatively large, and comparable to {\it e.g.} nucleon-nucleus cross section, then the Earth's atmosphere and overburden 
leads to a quick slow-down of DM particles \cite{Hooper:2018bfw,DeLuca:2018mzn, Neufeld:2018slx}. As a result, at the deep location of the most sensitive detectors, the DM velocity may be close to a thermal one, 
which is far smaller than the threshold velocities required for scattering. 
Thus, strongly-interacting DM (or fraction of DM) may evade being detected despite a possibly large scattering rate. 

Another well motivated class of DM are inelastic DM models where the DM possesses purely off-diagonal couplings at tree-level such that DM scattering off SM particles requires a transition to a higher excited state \cite{TuckerSmith:2001hy,Feldstein:2010su,Pospelov:2013nea,Bramante:2016rdh}. If the relative nucleus-DM kinetic energy is below the mass defect, the scattering is almost absent leading to much reduced sensitivity. Thus, in both these scenarios the kinetic energy deposited on the SM target might be far lower than regular DD leading to weakened experimental reach. In this paper, we propose a DD strategy that instead supplies energy to DM, to search for these well motivated classes of DM models.

 Constructing such an exothermic detector requires an energy source that can afford large exposures, {\em e.g.} $\mathcal{O}$(Avogadro number $\times$ year). One candidate for such a source are metastable nuclear isomers. These are typically high spin excited states of ground state nuclei which have long lifetimes due to their decay being governed by angular momentum selection rules. 
 But, scattering can violate this selection rule: a projectile impinging on the target can carry angular momentum from the state, leading to a prompt de-excitation of the nucleus. To illustrate this idea, consider, for example, an $L$-th multipole electric transition between two nuclear states. The transition 
 amplitude has a characteristic $\sim(k R_N)^L$ dependence, where $k$ is the momentum exchange in the transition, and $R_N$ is a typical nuclear size. For an on-shell photon, and high enough $L$, this results in a huge suppression 
 of the amplitude and longevity of excited states, as $k_\gamma = \Delta E_N$. For DM scattering, however, the momentum exchanged in the collision has a 
 completely different scaling, $\sim (\mu \Delta E_N)^{1/2}= 
 \Delta E_N(\mu/ \Delta E_N)^{1/2}$, where $\mu$ is the 
 reduced mass of the system. For weak scale and heavier DM, $\mu$ is 
 $\sim 6$ orders of magnitude larger than $\Delta E_N$, so that the transition amplitude can be enhanced 
 by an enormous amount. 
 
 Once the exothermic scattering occurs, the nuclear de-excitation energy is shared between nucleus and DM particle. (At some point, nuclear isomers were entertained as a possible energy storage, with energy extracted on demand using external energetic particles, but the process was not found to be energy efficient.) We propose to use the nuclear de-excitation process as an exothermic DM detector to search for slow or inelastic DM collisions. If the ambient DM triggers the decay of a metastable isomer and gets kicked in the process, we can look for the following processes: the de-excitation of the target nucleus to its ground state, subsequent decay of excited DM in detector volume, or a re-scatter of excited DM in a conventional DM DD experiment.

These ideas are explored in detail in what follows. In Section.~\ref{kinematics} we explore the general kinematics of DM scatter. Section.~\ref{meta} presents salient features of different metastable isomers and the various detection strategies. This is followed by Section.~\ref{rate} which deals with the rate for these processes given a reference cross-section as well as form-factors that capture nuclear scattering matrix elements. Section. \ref{sidm} discusses DM that interacts strongly with nuclei, and sets new limits and projections for the future and a similar exercise is carried out for inelastic DM in Section.~\ref{idm}. Finally we conclude in Section.~\ref{conclusion}

\section{Kinematics}
\label{kinematics}
To explore the full range of phenomenological possibilities discussed above, we consider the following kinematic setup. The DM particles come and scatter with the nucleus, de-exciting the nucleus. The energy released can go both into the kinetic energy of the DM and could potentially excite it to a higher inelastic DM state. Accordingly, we call the decrease in the internal energy of the nucleus to be $\Delta E_N$ and any increase in the internal energy of the DM to be $\Delta E_{\chi}$. The energy $\Delta E = \Delta E_N-\Delta E_\chi$ is the available kinetic energy that can be shared between the nucleus and the DM. 

If the DM with mass $M_{\chi}$ and with momentum $\bm{q_i}$ scatters to transfer momentum $\bm{q}=\bm{q_f}-\bm{q_i}$ and leaves with final momentum $\bm{q_f}$, from momentum conservation, the transferred momentum in the nucleus (mass $m_N$) is also $q$ and hence the recoil energy can be written as,
\begin{equation}
E_r=\frac{q^2}{2m_N}
\end{equation}
Energy conservation yields
\begin{equation}
\frac{q_i^2}{2M_{\chi}}=E_r - \Delta E +\delta M_{\chi} + \frac{q_f^2}{2M_{\chi}}.
\end{equation}
Defining the reduced mass, 
\begin{equation}
\mu_{\chi N} = \frac{M_{\chi} m_N}{M_{\chi}+m_N},
\end{equation}
and eliminating $q_f$ in the previous equation, we get,
\begin{equation}
\frac{q^2}{2\mu_{\chi N}}-\frac{q q_i \cos \beta}{M_{\chi}}-\Delta E =0.
\end{equation}
Thus given a DM initial velocity $v$, there is a range of momentum transfers given by
\begin{equation}
q^2_{\text{min/max}}(v)=\mu^2_{\chi N} v^2\left[1\mp \sqrt{1+\frac{2\Delta E}{\mu_{\chi N} v^2}}\right]^2.
\label{qminmax}
\end{equation}

In the limit of vanishing DM velocity, the momentum transfer is fully defined
in terms of masses and mass defects, 
\begin{equation}
q^2_{\text{min/max}}(v)\to q^2_0 = 2\mu_{\chi N}\Delta E.
\label{q0}
\end{equation}

As we will see below, the scattering of the DM can efficiently de-excite the nucleus only when it can carry momenta $\gtrapprox$ 100 MeV, {\it i.e.} be comparable to or larger than the inverse nuclear size. We thus see that this technique is most useful for DM candidates whose masses are $\gtrapprox 100$ GeV scattering off heavy $\sim 100$ GeV nuclear isomers. 

\section{Metastable nuclear isomers and possible DM search applications}
\label{meta}
In this section, we review the properties and abundances of metastable nuclear isomers that we believe are the most promising for DM searches. The signal of the DM scattering would either be the direct observation of the nuclear decay itself, or the observation of the possible re-scatters or decay of the excited DM states produced in the scattering event in a nearby DM detector. While the latter possibility can work for all isomers, the observability of the former process is isomer dependent. In the discussion below, we describe possible observational strategies for each isomer. 

Metastable nuclear isomers are higher excited states of nuclei 
which could undergo a gamma decay, but are comparatively 
long-lived typically due to high spin stabilization. 
Selection rules in quantum mechanics put parities and angular momenta of initial and final states $J_i$ and $J_f$ 
in correspondence with the multipole order of the photon transition and its magnetic/electric property (see {\em e.g.} \cite{Berestetsky:1982aq}). 

Since the wavelength of an outgoing gamma is much larger than the characteristic nuclear size, and thus any length scale associated with a given multipole transition, the multipole expansion works well, and the lowest multipoles are usually the most important.

\begin{figure}[htpb]
\centering
\includegraphics[width=1\columnwidth,keepaspectratio]{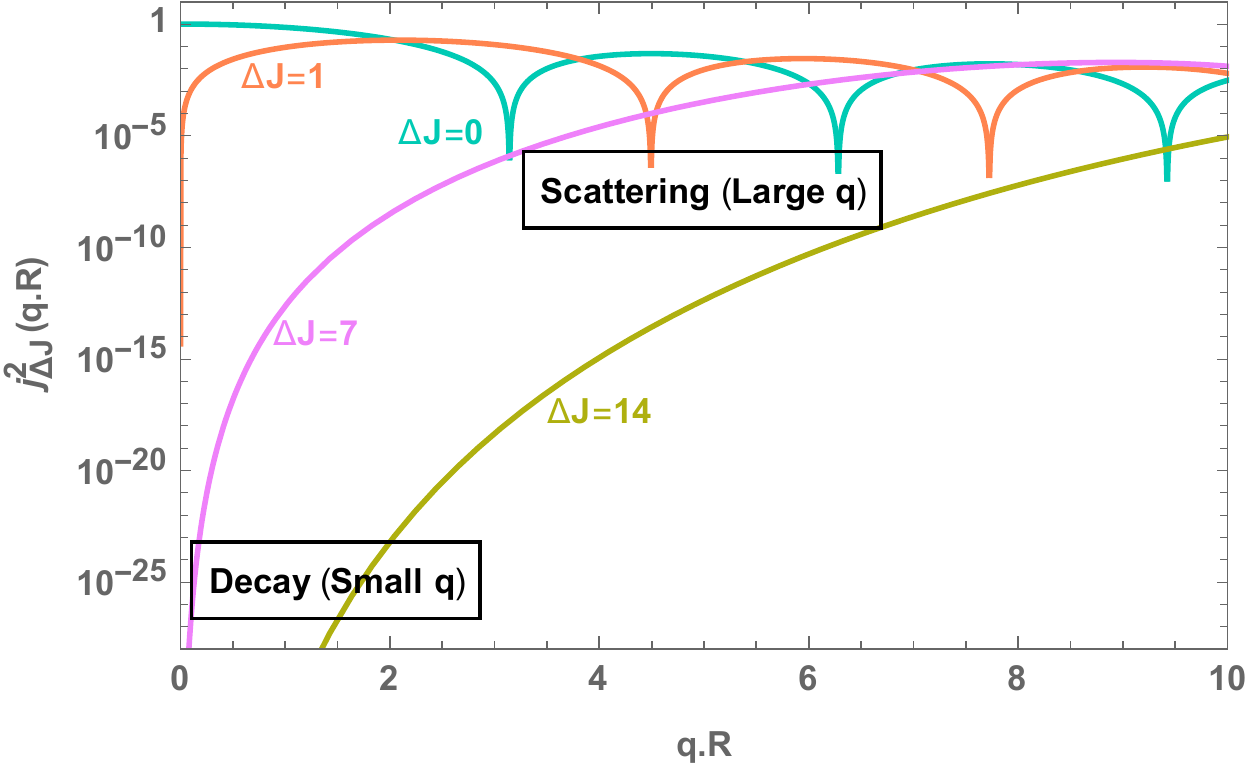}
\caption{Form Factors for different $\Delta J$. For small q.R relevant to $\gamma$ decay, there is severe suppression for large $\Delta J$. If the typical momentum required for scattering is much larger, the form-factor suppression is ameliorated.}
\label{fig0}
\end{figure}

As already being alluded to in the Introduction, the transition matrix elements are suppressed by powers of $(q.R)^L$, where $q$ is the outgoing photon momentum and $R$ is the size of the nucleus. While a typical gamma decay with $L\leq 2$ happens within a picosecond time scale, suppression for large $L$ can increase the lifetime to much larger timescales.
The {\em minimum} $L$ is given by $|J_f-J_i|$ (or sometimes by $|J_f-J_i|+1$, depending on 
electric/magnetic type of transition and matching of parities). This suppression is shown via the form factor (defined formally later) in Fig.~\ref{fig0} for different $\Delta J$. As seen in Table~\ref{table1} lifetimes of metastable isomers can be $\mathcal{O}({\rm min})$ or even $\mathcal{O}({\rm year})$. The nucleus most stable against radioactive decay is an isomeric state of $^{180}$Ta that has not been observed to decay and only a lower limit of $\tau >10^{16}$ year is known.

While metastable nuclei are resistant to gamma decay due to multipole suppression, scattering does not suffer from the same suppression in the large momentum ($q R \sim 1 \implies q \gtrapprox$ 100 MeV) exchange regime. In this regime, the multipole expansion of the outgoing DM wave receives contributions from many angular momentum modes, enabling the de-excitation transition. This can be seen quantitatively in Fig.~\ref{fig0} where the form-factor for different $\Delta J$ are no longer suppressed for larger momentum exchange. Many standard model projectiles \cite{karamian1999fast,hartouni2008theoretical,karamian2011cross} have in fact been employed to induce down-scattering of isomers, and extract the excess energy in the process. We are going to show that there are classes of DM particles 
that indeed lift this momentum suppression, and their direct detection scenarios 
would greatly benefit from the  scattering off metastable isomers. We now list nuclear isomer candidates and their properties.

\subsection{$^{180m}$Ta}
$^{180m}$Ta ($J=9^-$) has never been observed to decay while the corresponding ground state ($J=1^+$) decays with an 8 hour lifetime. This is because of the highly suppressed $E_7$ transition to the only other excited state ($J=2^+$). Thus, there is a significant abundance of this isomer in nature - it occurs with a yield of $0.011\%$ in naturally occurring tantalum. Further refinement has been carried out for making highly enriched tablets for decay studies. The effects of DM can be observed in $^{180m}$Ta either by monitoring a radio-pure sample wherein the down-scattering event triggers the decay of the ground state within 8 hours. Alternatively, large quantities of tantalum can be placed near a conventional DM detector, wherein the DM kicked or excited in this process can either re-scatter or decay within that detector.  

Null results from $^{180m}$Ta decay experiments \cite{Lehnert:2016iku} could in principle be already used to set limits on DM scattering. However in the above work, only $\beta$ and $\epsilon$ decay were considered. From private communications with B. Lehnert, the limit on half-life for the isomeric decays can be obtained with $\tau > 10^{14}$ years. In an accompanying paper \cite{forthcoming} we convert this into current limits on strongly interacting DM. The SM $^{180m}$Ta lifetime is estimated to be $\tau \sim 10^{17}$   years\cite{Ejiri:2017dro}. If a faster lifetime is experimentally observed, its connection to strongly interacting DM can be established by looking at the depth dependence of the lifetime. If the predicted standard model lifetime is experimentally observed, the experiment is no longer background free. However, progress can still be made by searching for the re-scattering of the DM in a well-shielded conventional DM detector. In spirit, this is similar to a light shining through walls experiment.  We consider a gram-year exposure with $\mathcal{O}(1)$ efficiency for subsequent detection of DM for such a setup. 

\subsection{ $^{177{\rm m}}$Lu}
$^{177{\rm m}}$Lu is $970$ keV above the ground state and has a half-life of $160$ days. It is a $0.1\%$ contaminant in $^{177}$Lu (with half-life $\sim$ 6 days) which is used in cancer treatment. The thermal neutron absorption cross-section on $^{176}$Lu to produce $^{177}$Lu and $^{177{\rm m}}$Lu are $2090$ barn and $2.8$ barn, hence the $0.1\%$ contamination. The long life of the isomer leads to medical waste containing most of the $^{177{\rm m}}$Lu intact, and it is typically shipped to nuclear waste facilities after usage. We assume $1$ mg $^{177{\rm m}}$Lu can be procured either from this medical waste or from dedicated production. Since the source is hot, only secondary detection of excited DM (either through decay or re-scatter) will be considered. 

\subsection{ $^{137{\rm m}}$Ba}

 $^{137{\rm m}}$Ba is a $661$ keV isomer but has a half-life of only 2.55 min. However, it is produced in the decay chain of $^{137}$Cs which exists in extremely high quantities in nuclear waste. While conducting any precision study on the isomer itself is futile in this case, a scheme with subsequent re-scattering or excited DM decay is possible. In particular there are some physics experiments that happen in the vicinity of nuclear waste facilities, which can possess large acceptance to DM that is re-scattered. For investigating the sensitivity reach, we will assume $100$kg $\times 2.55$ min, i.e. the experiment has exposure to 100 kg equivalent of $^{137{\rm m}}$Ba which lives for $2.55$ min, which were originally produced from 100 kg equivalent of $^{137}$Cs.
 
\subsection{ $^{178{\rm m}}$Hf}
$^{178{\rm m}}$Hf ($J=16^+$) is a $31$ year half-life isomer, with $2.46$ MeV energy excess over the ground state, which is the largest  among all long-lived isomers. The stability of the ground-state and the short life of the isomer makes it difficult to do a tantalum-like analysis where the ground state produced in the scatter could be observed to decay. Instead, we again rely on the possibility of the DM either re-scattering or de-exciting in a nearby detector. Another exotic possibility is the prospect of scattering to intermediate nuclear states not in the usual decay chain which produce gamma rays at higher energies than what is observed. About $1\mu$g of isomeric Hf has survived from experiments that explored the possibility of this isomer as an energy storage device. We will assume $1\mu$g$\times$year as the exposure for this isomer. 
 
%\subsection{ $^{108{\rm m}}$Ag}
{
\begin{table*}[ht]
\begin{center}
% \color{white}
 \begin{tabular}{||c | c | c | c | c | c | c | c | c||} 
 
 \hline

 Isomer & $\Delta E_N^{\text{max}}$ & levels & Half-life & Source & Amount & Signal &Hindrance ($F_\gamma$)\\ [0.5ex] 
 \hline\hline
 $^{180 {\rm m}}$Ta & 77 keV & 2 &$>10^{16}$ y & Natural & 0.3 gram year&Ground State Decay / Secondary & 0.16\cite{Ejiri:2017dro} \\
 \hline
% $^{108{\rm m}}$Ag & 109 keV & 2 & 438 y & & & & \\
%\hline
 $^{137{\rm m}}$Ba & 661 keV & 2 & 2.55 min & Nuclear Waste & 0.5 gram year &Secondary & 1 \\
 \hline
 $^{177{\rm m}}$Lu & 970 keV & 27 & 160 d & Medical Waste &1 mg year & Secondary & 0.17\footnote{\label{note1}Hindrance factors for Lu and Hf derived from the observed half-lifes.}\\
 \hline
 $^{178{\rm m}}$Hf & 2.4 MeV & 110 & 31 y & Old experiments & 1 $\mu$g year &$\gamma$ end-point / Secondary & $0.29^\text{\ref{note1}}$ \\
\hline

\end{tabular}
\end{center}
\caption{Isomers considered in this work are tabulated. The energy of the metastable state, the number of levels between the isomeric state and the ground state and the half-lifes of the isomeric state are given. Also tabulated are the typical exposure for each isomer used in projections. Possible trigger signals for isomer scattering are listed. Finally the hindrance factor used to calculate transitions/scattering cross-sections as used in Eqn.(\ref{eq:hindrance}) are given.}
\label{table1}
\end{table*}
}

\section{Rate}
\label{rate}
The counting rate of an experiment containing $N_T$ target nuclei to a particular final state
 is given by,
\begin{align}
R=N_T \frac{\rho_{\chi}}{M_{\chi}} \int_{q^2_\text{min}(v)}^{q^2_\text{max}(v)} dq^2\langle v \frac{d\sigma}{dq^2}\rangle.
\end{align}
For this paper, $N_T$ counts the excited nuclear states, 
that DM scattering can de-excite, making $J_i\rightarrow J_f$
nuclear transition.

The DM velocity distribution could either be given by a galactic Maxwell-Boltzmann-type distribution boosted with the particular velocity of the earth and solar system, 
or by a truly thermal distribution inside the Earth with a small terminal velocity component for strongly interacting DM (see below).

Scattering from the isomeric to the ground state provides the largest $\Delta E_N$, $\Delta E_N^{\rm max}=E_N^{\rm iso}$, the energy of the metastable isomer state when the ground state energy is taken to be zero. There are often states between the metastable isomer and ground states that can provide additional reach at $0\le E_N^{\rm max} - E_N^{f}= \Delta E_N\le E_N^{\rm max}$. These isomeric states are summed over with the appropriate $J_f$ and $\Delta K$ quantum numbers to get the full scattering rate. These states can be arranged in $K$ branches (which will be introduced next), with each branch containing states $J=K,K+1,K+2...$ with energies typically increasing. The $J$, $K$ quantum numbers as well as $E_N^f$ for the various nuclear candidates are given in Appendix.~\ref{append}.

Summing over all final states $f$, $R$ can be written as
\begin{align}
R=N_T \frac{\rho_{\chi}}{M_{\chi}} \int d^3 v f(v) v \sum_f \int_{q^2_\text{min}(v,E_N^f)}^{q^2_\text{max}(v,E_N^{f})} dq^2 \frac{d\sigma_N}{dq^2}\mathcal{S}_f(\bm{q}),
\label{eqn:Rate}
\end{align}
where $\sigma_N$ is the reference cross-section, where most of the $q$-dependence is factored into $\mathcal{S}_f(\bm{q})$. $\mathcal{S}_f(\bm{q})$ parametrizes the $q$-dependence of the nuclear transitional matrix element squared, and we will refer 
to it as "nuclear form factor" for convenience.  

\subsection{Nuclear inelastic form factor}
 In this subsection, we will consider $\mathcal{S}_f(\bm{q})$ following the 
 approach of Ref. \cite{Engel:1999kv}. Furthemore, we shall assume that the 
 main form of the DM-nucleus interaction is 
 scalar-scalar, $\bar \chi \chi \bar NN$, or vector-current type,
  $\bar \chi \gamma_\mu \chi \bar N \gamma_\mu N$, which are identical 
  in the non-relativistic limit. Very similar treatment will apply to the spin dependent case as well. The quantity $\mathcal{S}_f(\bm{q})$ originates 
  from $ | \langle J_f| \sum_{i} \exp(i \bm{q}\bm{r}_i)|J_i \rangle |^2$, where 
  the sum is taken over the individual nucleons.  Calculation of this complicated 
  matrix element cannot be performed in full generality without a proper knowledge of 
  full multi-nucleon wave functions of the initial and the final states. We take an approach of estimating this matrix element in a Weisskopf-style estimate, 
  after accounting 
  for $q$ dependence and known conservation laws. Where possible we connect these 
  matrix elements to the one known via gamma
  transition. 

Expanding this quantity in spherical harmonics, we get
\begin{equation}
\mathcal{S}_f(\bm{q})= \sum_{L}|\langle J_f||\sum_i j_L(q r_i) Y_{LM}||J_i \rangle |^2,
\end{equation}
%{\bf check equations for states with opposite parity ??}
where summation goes over the nucleons participating in the transition. 

Assuming that the transition density is highly peaked at the surface, Ref. \cite{Engel:1999kv} (as this is a very good approximation for higher multipole transitions \cite{sandor1993shape}), and is mediated by one or a few valence nucleons, 
we arrive at a following estimate:
\begin{equation}
\mathcal{S}_f(q)\simeq \sum_L j_L^2(qR) \epsilon_H,
\end{equation}
where $R$ is the nuclear radius, and $\epsilon_H$ is an additional ``hindrance factor'', on top of a regular angular momentum suppression \cite{Ejiri:2017dro}. In a proper nuclear calculation, $\epsilon_H$ would reflect the overlaps of the wave functions inside the 
transitional matrix element beyond the angular momentum factor taken into account by
$j_L^2(qR)$. 

Due to angular momentum conservation, the sum over $L$ only needs to be taken in the range $|J_i-J_f|\le L \le J_i+J_f$ keeping only the appropriate parity terms. Hindrance factors arise due to the so-called $K$-quantum number selection rules in deformed nuclei. The $K$-quantum number captures the misalignment of the rotation and symmetry axes and if the multipolarity of the transition $L$ is smaller than than $\Delta K$, this leads to a suppression factor
\begin{equation}
\epsilon_H (L,\Delta K)=\frac{T_\gamma(\text{naive})}{T_\gamma}=F_\gamma ^{\Delta K-L} \quad \text{if}~L < \Delta K
\label{eqn:hindrance1}
\end{equation}
Here, $T_\gamma$ is the measured (predicted in the case of tantalum) lifetime and $T_\gamma(\text{naive})$ is the lifetime predicted from naive Weisskopf estimates. 

For barium decays the naive Weisskopf estimate captures the observed decay rate extremely well. There are two states the metastable isomer can scatter to\cite{nudat20112}. For hafnium and lutetium, we estimate $F_\gamma$ from the observed decay rate and tabulated in Table.~\ref{table1}. They also have a plethora of states to scatter to, with different $K$-quantum numbers. The full list of energy levels and their $J$ and $K$-quantum numbers can be obtained from \cite{nudat20112} and are tabulated in Appendix.~\ref{append}. We use the same $F_\gamma$ for different $\Delta K$ transitions. At high momentum exchange as is relevant to scattering, then, scattering can proceed through $L\ge \Delta K$ avoiding the suppression in Eqn.(\ref{eqn:hindrance1}). As a result the total scattering rate is only weakly dependent on the hindrance factor.     

Since tantalum has never been observed to decay, the above procedure does not apply. In \cite{Ejiri:2017dro}, extra penalty factors independent of $\Delta K$ are prescribed,
\begin{equation}
\epsilon_H^{\rm Ta}(J_i,L,\Delta K)  = M_0(EL)^2 (F_\gamma ^{\Delta K -L})^2
\label{eq:hindrance}
\end{equation}
where  $M_0(EL)=0.35$ and $F_\gamma=0.16$.

For tantalum, this procedure is only an order of magnitude estimate at best: after all, an $E_7$ transition has never been observed in nature to enable robust extrapolation. We use this for our estimates to both states below the metastable isomer. Coulomb scattering with a scheme similar to \cite{sandor1993shape} could be very interesting in its own right to estimate this form factor more accurately and also to get a better understanding of DM interaction rate.

\begin{comment}

\subsection{Intermediate states}

For $^{180 {\rm m}}$Ta, the intermediate (and ground) states below the isomer are in $\{K,J,E_N^{\rm inter}{\rm [keV]}\}$ notation, $\{1,1^+,0\},\{1,2^+,39.54\}$ and the isomer state is $\{9,9^-,77.2\}$

For $^{137 {\rm m}}$Ba, the $K$ states are not relevant since there is no suppression, the intermediate (and ground) states below the isomer are 
$\{\frac{3}{2}^+,0\}, \{\frac{1}{2}^+,283.54\}$ with the isomeric state at $\{\frac{11}{2}^-,661.659\}$.

For $^{177{\rm m}}$Lu, there are 27 states we consider below the isomer state, and are classified into 6 $K$-branches: $\{\frac{1}{2},\frac{3}{2},\frac{5}{2},\frac{7}{2},\frac{9}{2},\frac{13}{2}\}$. Their specific energy levels and $J$ can be found in \cite{nudat20112}. The isomer state itself has $K=\frac{23}{2}$.

$^{178{\rm m}}$Hf has a whopping 110 levels to scatter to. They belong to the $K$ branches, $\{0, 1, 2, 3, 4, 5, 6, 8, 10\}$ and the entire set can be found in \cite{nudat20112} with the metastable isomer itself at $K=16$. 

Wherever relevant, $F_\gamma$ is evaluated only for the observed $\Delta K$ transitions and the same $F_\gamma$ is used for evaluating all other $\Delta K$ transitions.
\end{comment}

\section{Strongly Interacting DM}
\label{sidm}
For DM that interacts strongly with nuclei, limits from traditional experiments in underground laboratories are not relevant. This is because the DM slows down considerably and does not possess enough kinetic energy to scatter off the target nuclei. This parameter space is constrained by surface runs of some experiments and balloon based cosmic ray detectors. While this parameter space is constrained if all of the DM is strongly interacting, due to the small exposure of these experiments, there are no constraints on strongly interacting DM if it only makes up a somewhat smaller sub-component of the DM. Moreover, models with exotic heavy QCD-charged remnants that 
form heavy compact bound states and comprise the bulk of DM, also predict small concentrations of particles that have strong interactions with nucleons~\cite{DeLuca:2018mzn}. 

The phenomenology of this kind of DM is as follows: the DM undergoes a series of scatterings with nuclei, that slows down from the initial $v\sim O(10^{-3})c$ velocity, 
leading to eventual thermalization.  Upon thermalization, the DM acquires a thermal randomly oriented velocity. In the earth's gravity and if there is no DM binding to nuclei, the DM particles slowly drift downward. Since the downward drift velocity of the DM is much smaller than the ambient virial velocity in the galaxy, there is a pile up (or ``traffic jam'') of the DM as it moves through the ground, leading to a very significant local density enhancement compared to DM density beyond Earth's atmosphere. Thus, in an underground experiment, there is an enhanced density of slow DM, inaccessible to all 
DD experiments looking for elastic scattering. Since this DM is heavy, it can scatter off the nuclear isomers discussed above, producing measurable signals either directly in the process of nuclear de-excitation, 
or in the subsequent elastic collision - now with much larger energy. In the following, we compute these effects and estimate the reach for an isomer-based concept experiment. 

 %There are considerable uncertainties and conflicting claims about whether this region is adequately covered by a combination of balloon and ground based experiments. Regardless, for subcomponent DM and spin-dependent DM, unambiguously, there are gaps between the reach for balloon experiments and where the ground based experiments start becoming sensitive.

\subsection{The DM Traffic Jam}

To estimate the density enhancement in the DM traffic jam, we begin by first estimating the terminal velocity with which the DM sinks through the ground. The density enhancement then follows from flux conservation. 

\begin{figure*}[htpb]
\centering
\includegraphics[width=0.98\columnwidth,keepaspectratio]{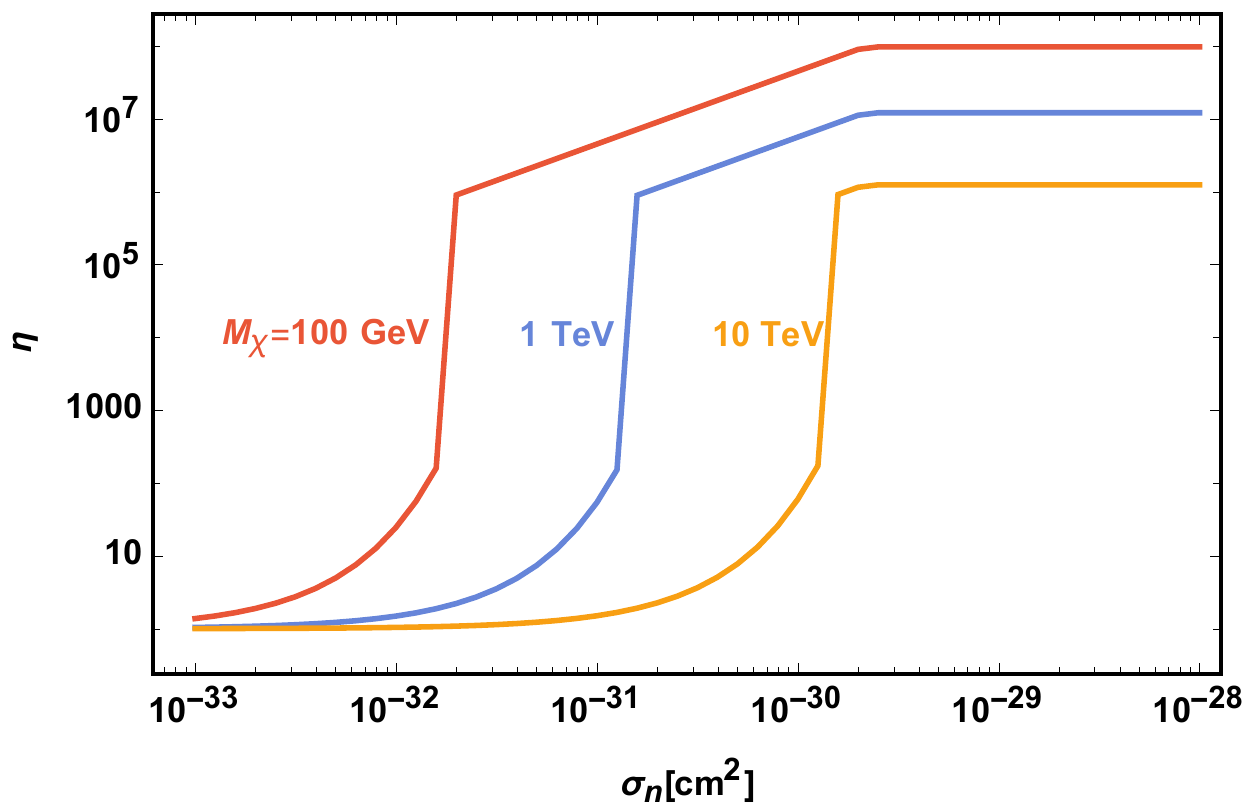}
\includegraphics[width=0.98\columnwidth,keepaspectratio]{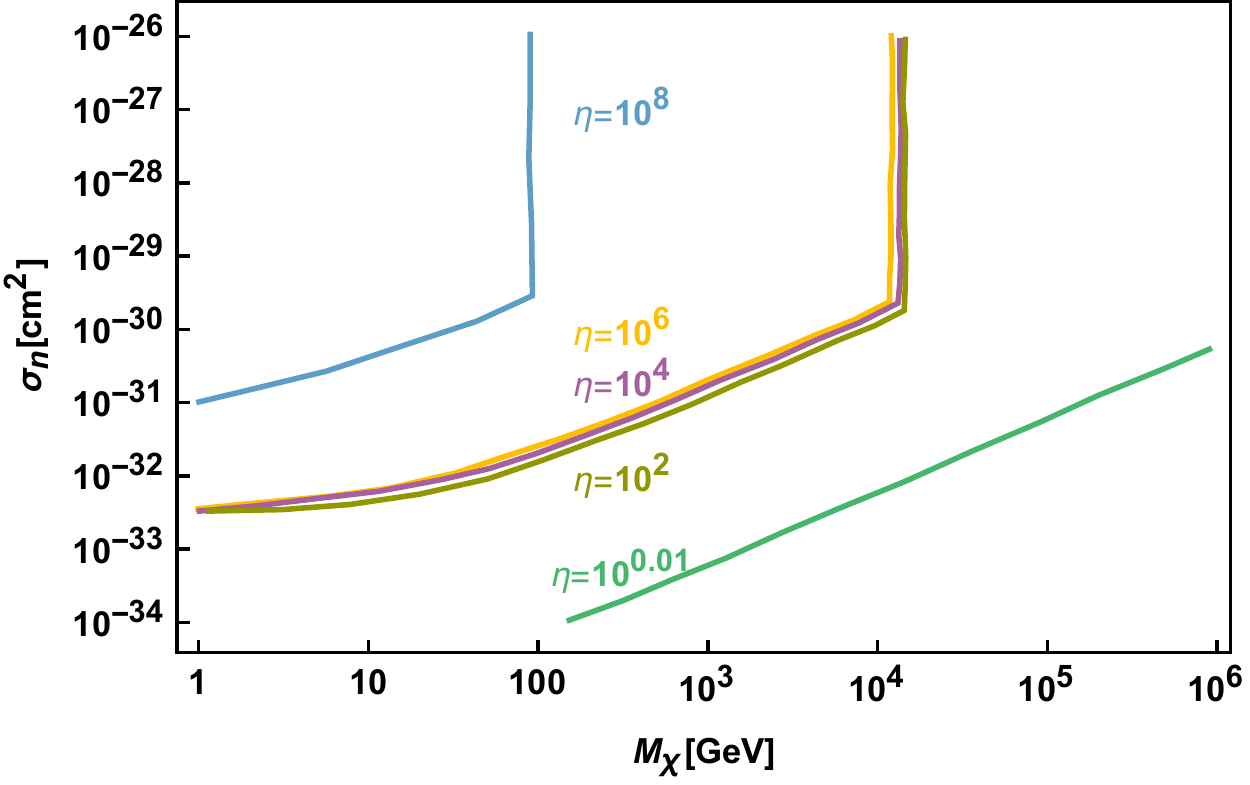}
\caption{{\bf Left:} Density enhancement $\eta$, 300 meters below the surface, for 3 different DM masses, $M_{\chi}=100$GeV, $M_{\chi}=1$TeV, and $M_{\chi}=10$TeV as a function of reference nucleon cross-section $\sigma_n$. {\bf Right:} Contours of constant density enhancement $\eta$ in the $\sigma_n$ vs $M_{\chi}$ plane.}
\label{figenhance}
\end{figure*}

We work in the limit where the DM interacts sufficiently strongly with nuclei so that it thermalizes when it goes underground. This is the range of parameters that is of most interest, since the scattering of DM is otherwise constrained by low threshold detectors such as CRESST. Thermalization is of course progressively harder at heavier masses since several collisions are necessary for the DM to thermalize with the rock. To avoid rather strong constraints on anomalous isotopic abundances, we will assume that the strongly interacting DM has repulsive
interaction with nuclei. 

To perform an estimate of the density enhancement, 
we need a coherent (transport) scattering cross-section $\sigma_t$ of DM with nuclei of atomic mass $A$. We notice that in principle, there are two main 
regimes for such a scattering cross section. The first regime can be 
achieved when the perturbative treatment is possible. Then, given 
the input cross section on an individual nucleon, the overall 
elastic cross section on the nucleus could be described as 
$\sigma_{el} =A^2 \sigma_n\mu^2(m_A,m_\chi)/m_p^2 $, 
which reduces to $A^4\sigma_n$ at $M_\chi\gg m_A$. On the hand, if we keep 
increasing $\sigma_n$ this scaling with $A$ breaks down. Describing the DM-nucleus 
potential as a square barrier, we observed that 
the strong interaction limit corresponds 
to $R_A \kappa \gg 1$, where $\kappa$ is the virtual momentum inside the 
barrier \cite{Landau:1991wop}, and the elastic cross section is expected to be 
$4\pi R_A^2$. For the slow-down process, we need a transport cross section, 
and we assume it to be on the same order of magnitude as the elastic one. 
Thus, we choose the following ansatz for the $\chi$-nucleus transport cross 
section,
 \begin{equation}
\sigma_t=\text{Min}(A^4 \sigma_n, 4\pi R_A^2).
\label{saturation}
\end{equation}

After DM is fully thermalized, it is not stationary, but continues slowly sinking towards the center of the Earth due to the Earth's gravitational field. The average terminal downward velocity in any medium is given by \cite{landau1981course} 
\begin{equation}
v_\text{term}=\frac{3 M_\chi g T}{m_{\rm gas}^2 n \langle \sigma_t v_\text{th}^3 \rangle}
\label{terminal}
\end{equation}
where $M_\chi$ is the DM mass, $m_{\rm gas}$ is the mass of gas particle, n is the number density of gas particles,
$\sigma_t$ is the transport cross section, $v_\text{th}$ the thermal velocity of gas particles (for solids, velocity due to vibrational motion)\footnote{This effect was discussed in \cite{DeLuca:2018mzn}. However, their estimate differs from the calculations of  \cite{landau1981course}. Moreover, \cite{DeLuca:2018mzn} did not account for the saturation of the DM nucleon scattering cross-section at large $A$ and did not use the correct reduced mass in the collision between DM and nuclei.}.

This terminal velocity $v_\text{term}$ is lower than the initial (galactic) DM velocity, leading to the DM pile up and a resulting density enhancement. From flux conservation, the density enhancement is:
\begin{equation}
\eta=\frac{\rho_{\rm lab}}{\rho_{\rm ss}} = \frac{v_{\rm vir}}{v_{\rm term}} 
\end{equation}
where $\rho_{\rm lab}$ is the DM density at a location of an underground lab, $\rho_{\rm ss}$ is the solar system DM density, and $v_{\rm vir}$ is the local virial velocity of DM.

This density enhancement exists as long as the DM thermalizes with the rock. However, for heavy enough DM there are two additional effects that need to be taken into account. For large $m_\chi$ the thermalization requires more scattering, and there will eventually not be enough column depth in the rock to achieve thermal velocity at a given laboratory depth. Moreover, when the downward velocity of DM becomes smaller in magnitude than $v_{\rm term}$, the thermalization is not complete, as on average the vertical component of the DM velocity is larger than the terminal sinking velocity.  Both of these effects cut off the density enhancement for heavy DM, as shown in Fig.~\ref{figenhance} and discussed below. 

Many underground labs with developed DD program are located at depths exceeding 1 km. 
However, the precision experiments with metastable tantalum were performed in the Hades observatory, at a more shallow location. 
For our estimates, we take the Hades observatory to be 300 m below the surface. In our estimates, we take the density of soil/rock to be $\rho = \frac{3 \, \text{gm}}{\text{cm}^3}$, ambient temperature $T=300 \, \text{K}$, $m_{\rm gas} \sim A \times \, \text{GeV}$ and take $\text{A} \sim 30$ for rock. With these numbers, we plot the density enhancement $\eta$ for three different masses $M_{\chi}=100 \, {\rm GeV},1 \, {\rm TeV}, 10 \, {\rm TeV}$ in Fig.~\ref{figenhance} (Left). There are three distinct regimes at play. For small cross-sections, there is an exponential regime where the column density is not enough to slow DM particles down to the thermal velocity $v_{\rm th}$. As the downward velocity approaches the thermal velocity, the slow down is enhanced leading to a jump to $v_{\rm th}$. Next, for cross-sections where vertical velocity drops below $v_{\rm th}$, the additional column density leads to further slowing down, leading to a linear regime: the DM density 
enhancement is linearly proportional to the size of the elastic cross section. Finally, once
$v_{\rm term}$ is reached, there is no further slow down and a flat regime for the density enhancement is achieved.

Fig.~\ref{figenhance} (Right) shows contours of equal $\eta$ in the $\sigma_N$ vs $M_\chi$ plane. $\eta$ increases as a function of $\sigma_n$ till $\sigma_n \sim 10^{-30}~\text{cm}^2$ which corresponds to the saturated geometric cross-section in Eqn.(\ref{saturation}) and there is no further enhancement. As mass of DM, $M_\chi$ is dialed up, the terminal velocity increases linearly as in (\ref{terminal}), and as a result $\eta$ decreases linearly. However for large enough mass, the relevant column depth is not enough to thermalize and hence there is an exponential decrease in $\eta$ as a function of $M_\chi$. Thus, we conclude that the value of the 
enhancement factor is quite sensitive to particular details of 
the strongly-interacting DM model (mass, cross section), and can vary in a large range.

\subsection{Rate}
For DM that interacts strongly with nuclei, the relevant limit of Eqn.(\ref{qminmax}) is that of small initial velocity $v$, and $\delta_{M_{\chi}}=0$. Thus we get $q^2 \sim q^2_{0} \equiv 2\mu_{\chi N} \Delta E_N $ and 
\begin{equation}
q^2_{\text{min}}(q^2_{\text{max}}) = q^2_0 \mp 2 q_0 \mu v,
\end{equation}
where the second term is smaller than the first.

At this point, we would like to model a generic strong-interaction cross-section by exchange of a meson-like hadronic resonance; the differential cross section is given by,
 \begin{equation}
 \frac{d\sigma}{dq^2} = \frac{y_n^2 y_\chi^2}{8\pi(m_h^2+q^2)^2 v^2} \mathcal{S}_f(\bm{q_0}),
 \end{equation}
where $m_h$ stands for the mass of a typical strongly-interacting mediator ($\rho$, $\omega$, $\sigma$, $\pi$ etc mesons), $y_n$ and $y_\chi$ are the nucleon and DM Yukawa couplings.
After performing the $q^2$ integral from $q^2_{\text{min}}$ to $q^2_{\text{max}}$, we get,
\begin{equation}
\sigma v= \frac{q_0 \mu y_q^2 y_\chi^2}{2\pi (m_h^2+q_0^2)^2} \mathcal{S}_f(\bm{q_0}).
\end{equation}

Finally, the counting rate is given by,
\begin{align}
R=N_T \frac{\rho_{\chi}^{\rm local}}{M_{\chi}} \frac{q_0 \mu y_q^2 y_\chi^2}{2\pi 
(m_h^2+q_0^2)^2} \mathcal{S}_f(\bm{q_0}).
\end{align}

As often is the case for exothermic reactions, instead of depending on the DM flux, the counting rate will be uniquely sensitive to just the local DM density irrespective of its velocity. Using a sensible estimate of $m_h \sim \Lambda_{\rm QCD} \sim q_0 $, putting this in the previous formulae, and recognizing that combination ${y_q^2 y_\chi^2} \Lambda_{\rm QCD}^{-2} $ scales the same way as $ \sigma_n$, we find the following 
ansatz for the counting rate,
\begin{align}
R=N_T \frac{\rho_{\chi}^{\rm local}}{M_{\chi}} \text{Min}(\sigma_n \frac{\mu}{q_0} ,4\pi R_A^2)\mathcal{S}_f(\bm{q_0}).
\end{align}

Unlike exothermic scattering of inelastic DM, the scattering of strongly interacting DM on metastable nuclear states does not involve an energy barrier. We simply need a detectable signal from such de-excitation. Isomeric form of tantalum is adequate for this purpose, and we focus on it since it is a naturally occurring isomer enabling large exposure. DM that is accumulating in the underground laboratory, can scatter off $^{180m}$Ta to produce either the lower excited state or ground state of tantalum. Both of these are unstable and $\beta$-decay with much shorter time scales. This results in subsequent $\gamma$ radiation. These final state gamma quanta can be looked for to test this DM hypothesis. Isomeric decay of $^{180m}$Ta produces a near identical signature, albeit with an extra $\gamma$ line corresponding to the isomeric decay itself. Thus limits on the isomeric decay through limits on the subsequent $\gamma$ can be recast as limits on tantalum de-excitation caused by DM. Once a limit is set on tantalum decay rate, $\Gamma_{\rm lim}$, we can set a limit on $\sigma_n$ through,

\begin{align}
\sigma_n \le \Gamma_{\rm lim} \frac{M_{\chi}}{\rho_{\chi}^{\rm local}} \frac{q_0}{\mu} \frac{1}{\mathcal{S}_f(\bm{q_0})}
\end{align}

\begin{figure}[!htpb]
\centering
\includegraphics[width=0.98\columnwidth,keepaspectratio]{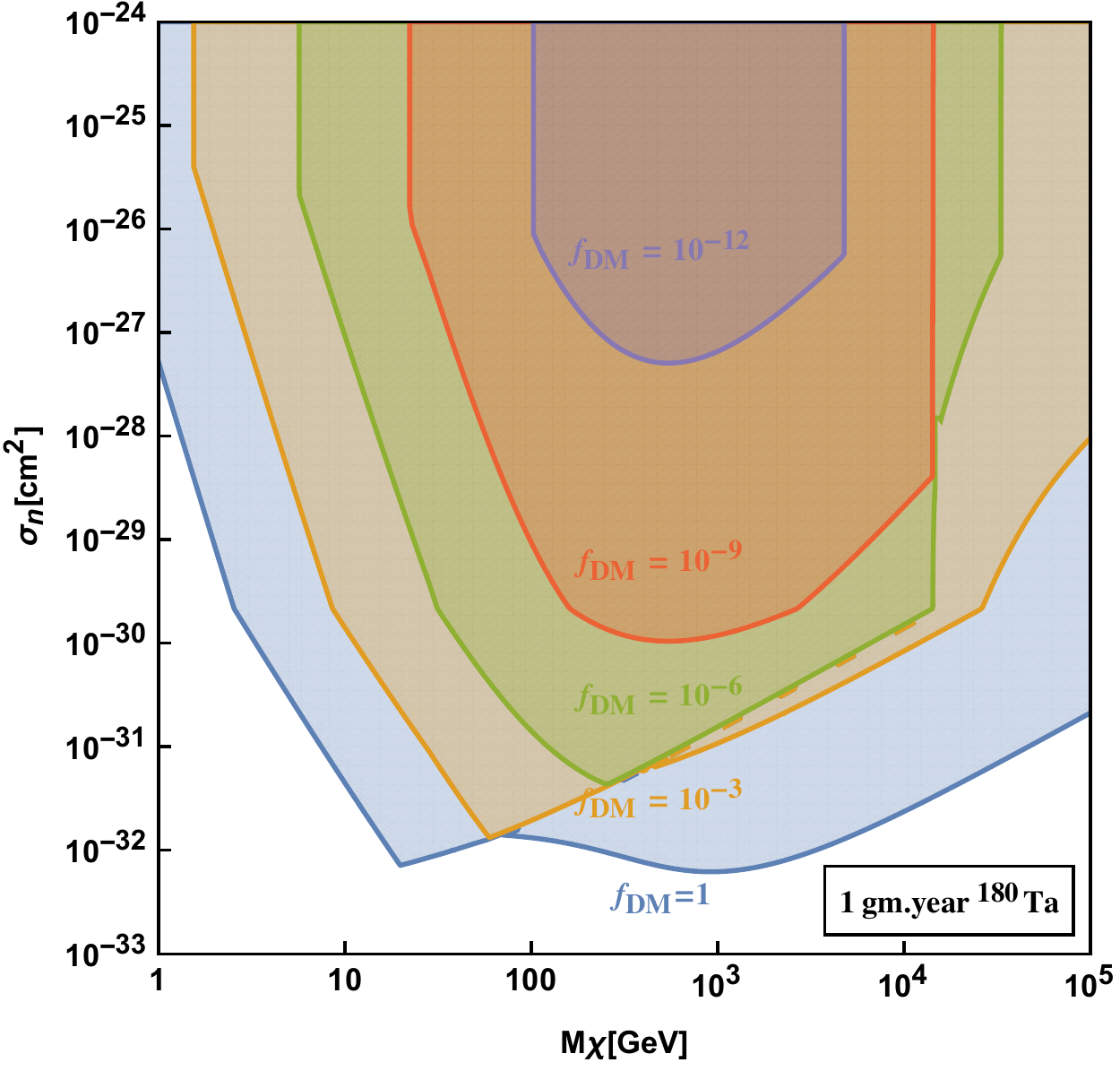}

\caption{
%{\bf Left:} 
Projections for limits on nucleon cross-section for DM that interacts strongly with nuclei that correspond to 1 gram year exposure for $^{180m}$Ta}
\label{Strongconst}
\end{figure}

In Fig.~\ref{Strongconst}, we make projections as a function of $\sigma_n$ and $M_\chi$ for a set-up that can set limits of $\tau > 10^{21}~ {\rm year}$ or equivalently an exposure of $1$ gram year. Such a setup can look for DM fractions $f_{\rm DM} \sim 10^{-6}$ in a wide mass range and fractions as small as $f_{\rm DM} \sim 10^{-12}$ in the $M_{\chi} \sim$ TeV range. Even with existing data (not shown), it should be feasible to set limits on sub-component DM that is strongly interacting with nuclei ($\sigma_n \sim 10^{-26} \text{cm}^2$) all the way down to a DM fraction of $f_{\rm DM}=10^{-4}$ in the wide range of mass $M_{\chi} \in \{20 \, \text{GeV}, 1 \, \text{TeV} \}$. In a narrower mass range, even fractions of $10^{-6}$ can be ruled out. We report these results along with the relevant experimental analysis in a companion paper~\cite{forthcoming}. 

Intriguingly, there is a strong depth dependence of DM fraction that is in a ``traffic-jam" and as a result, conducting the experiment at different depths could greatly help with signal-background discrimination, as well as 
potentially providing a smoking gun discovery signature. Of course, the quality of these limits
do depend in a crucial way on the value for the inelastic nuclear transition. Therefore, 
a dedicated nuclear theory evaluation of such transition is highly desirable.

\section{Inelastic DM}
\label{idm}
Inelastic DM is a blanket term to a plethora of models of DM that have purely off-diagonal interactions with SM at tree-level. These models were discussed originally in \cite{Hall:1997ah}
as a way of relaxing limits on sneutrino DM, and generalized in \cite{TuckerSmith:2001hy} to reconcile null results in germanium DD experiments with the DAMA excess that uses NaI crystals. While subsequent experiments with a heavier nucleus, Xe, were unable to confirm the DAMA excess, inelastic DM is still theoretically appealing. For example, by introducing a small energy splitting, the model reproduces the ``WIMP miracle'' ({\it i.e. } a natural explanation for the relic abundance of the DM) while eliminating limits from DD experiments. Inelastic scattering is also a natural expectation in composite DM models. For example, if the standard model interacts with the composite DM through a photon, the leading interaction would typically be through an inelastic dipole moment transition. 

Current limits on several inelastic matter models are summarized in \cite{Bramante:2016rdh}. Common to all these models is the mass-splitting $\delta M_{\chi}$. The maximum mass-splitting achievable in conventional detectors is $\delta M_{\chi} \le \frac{1}{2} \mu_{\chi n} v_{\chi}^2 $, i.e. the initial kinetic energy in the center-of-mass frame. The aim of this section is to explore the possibility of scattering off metastable nuclei enabling access to dark models where the splitting is bigger than the virial kinetic energy of the DM. With a splitting $\Delta E_N$, we can probe 
\begin{equation}
\delta M_{\chi} \le \frac{1}{2} \mu_{\chi n} v_{\chi}^2 +\Delta E_N
\label{maxsplit}
 \end{equation}
For this approach to constrain interesting parts of inelastic DM parameter space, the cross-sections that can be probed must be small enough so that the associated loop level elastic scattering cross-section is not ruled out by DD. The comparison between the inelastic and the loop level elastic scattering process is model dependent. We address it below for two well motivated examples of inelastic DM. 
 
In order to understand the reach of different metastable nuclei, let us start with a four-fermion operator, 
\begin{equation}
\mathcal{L} =G_F^D\bar{\chi}_2 \chi_1 \bar{N}{N}  +{\rm h.c}
\end{equation}
 %Without knowledge of the specifics of the UV theory it is impossible to calculate the corresponding elastic loop diagram so we will postpone that discussion to the next subsection.
Inelastic DM scatters off nuclear targets, down scatters the isomer to a lower excited/ ground state if there is enough energy available to scatter up to $\chi_2$. The corresponding observable is nuclear target dependent or DM model dependent. Generically we envision a mechanism to detect the excited state DM either through exothermic re-scatter inside a conventional DM or neutrino detector, or through a decay into SM particles. We assume the exposures justified in Section.~\ref{meta} and compute sensitivity reach for illustrative purposes only, postponing discussion of particular signatures to the specific models in the next subsections.

The differential cross-section is then given by
\begin{equation}
\frac{d\sigma}{dq^2} = \frac{\sigma_n}{v^2 \mu_{\chi n}^2} 
\label{diff}
\end{equation}
where $\sigma_n$, the reference DMDD nucleon cross-section is given by,
\begin{equation}
\sigma_n = \frac{(G_F^D)^2 \mu_{\chi n}^2}{8\pi}
\label{4fermi}
\end{equation}

\begin{figure}[t]
\centering
\includegraphics[width=1\columnwidth,keepaspectratio]{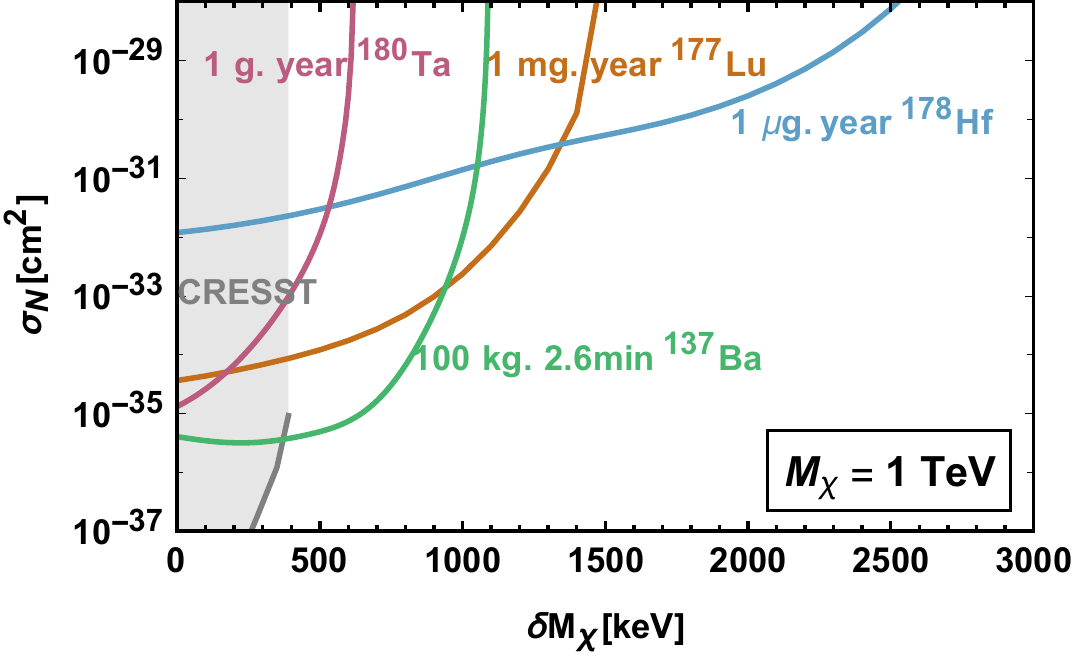}

\caption{DM-nucleon cross-section reach for different Isomeric nuclei for corresponding exposure. Also shown are "inelastic frontier" limits from CRESST. The sensitivity is derived assuming 3 detectable interactions per year. These limits are for illustrative purposes only. Limits on concrete models with detection signature are displayed in Fig.~\ref{fig2} and Fig.~\ref{fig3}}
\label{fig1}
\end{figure}

The reach for the cross-section in Eqn.(\ref{4fermi}) as a function of $\delta M_{\chi}$ is calculated using Eqn.(\ref{diff}) in Eqn.(\ref{eqn:Rate}) and is plotted in Fig.~\ref{fig1}. This figure shows the ultimate reach of detecting inelastic DM given realistic quantities of nuclear isomers, and assuming that three de-excitation events per year as the limiting rate. Therefore, importantly, for deriving these 
sensitivity curves we assume that the de-excitation events could be observed with high efficiency. (Such efficiency could approach ${\cal O}(1)$ if the decay length of the excited 
state for DM is commensurate with the linear size of the external detector that covers a significant part of solid angle, or if the 
de-excitation events can be observed {\em in situ}.)
For small $\delta M_{\chi}$ limits from conventional detectors are strongest purely from exposure considerations. Among these CRESST is dominant due to having the heaviest element element used in 
DD, tungsten ($^{187}$W). However these limits disappear altogether at $\delta M_{\chi}> 400$\,keV. Projections for limits from metastable nuclei extend well above that, with the maximum splitting given by Eqn.(\ref{maxsplit}). Potential limits from $^{178}$Hf are weak due to small mass of target but extend the farthest in $\delta M_{\chi}$. $^{180m}$Ta could have only nominal improvements in the splitting due its modest energy splitting ($77$ keV). $^{177}$Lu and $^{137}$Ba have theoretically a good reach in cross-section and should also be able to probe splittings above $1$ MeV. 

In practice, the lifetime and scattering rate of the excited state $\chi_2$ will determine the feasibility of this reach. These quantities, as well as the corresponding elastic loop-induced scattering cross-section are model-dependent, and will be considered for specific models next.

\subsection{Dark Photon Mediator}

\begin{figure}[htpb]
\centering
\includegraphics[width=1\columnwidth,keepaspectratio]{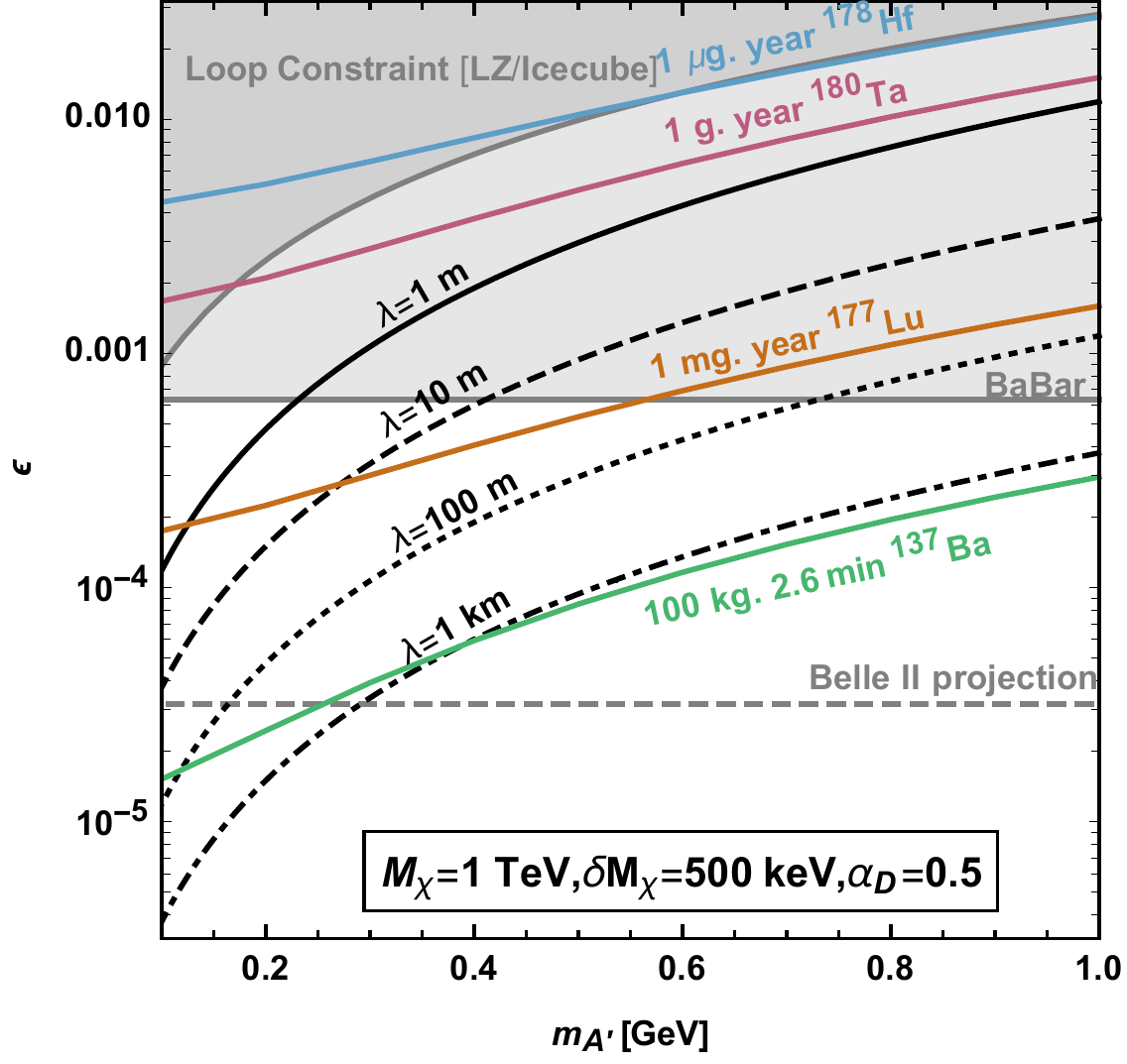}
\caption{Reach for different Isomeric nuclei for dark photon DM model. Limits from LZ/PANDAX and CRESST do not have any reach for $\delta M_{\chi} >450 \text{keV}$. Also shown in gray are loop constraints from LZ/Icecube and limits on the dark photon mediator from BABAR and projections for Belle II (dashed). Contours of constant mean-free-path for excited DM in a conventional DM detector are also marked. }
\label{fig2}
\end{figure}

We start with the terms in the Lagrangian relevant to the dark photon \cite{Izaguirre:2017bqb},
\begin{equation}
\mathcal{L} \supset i g_{\rm D} \overline{\chi}_2\gamma^{\mu} \chi_1 A'_\mu
 +{\rm h.c}+
\frac12 \epsilon F'_{\mu\nu}F^{\mu\nu}.
\end{equation}

The differential cross-section for inelastic scattering is given by \cite{Bramante:2016rdh},
\begin{equation}
\frac{d\sigma}{dq^2}=\frac{4 \pi \alpha \alpha_D\epsilon^2}{(m_{A'})^4 v^2}.
\end{equation}
where $m_{A'}$ is the mass of the dark photon, $\alpha = \frac{g^2}{4\pi}$ and $\alpha_D = \frac{g_D^2}{4\pi}$  . This can be substituted in Eqn.(\ref{eqn:Rate}) to obtain the event rate.

The elastic process at one loop level is given by \cite{Bramante:2016rdh} (see also \cite{Batell:2009vb}),
\begin{equation}
\sigma_{\text{n,loop}}=\frac{\alpha^2_D \alpha^2 \epsilon^4 m_n^4 f_q^2}{\pi (m_{A'})^6},
\label{elasticDF}
\end{equation}
for $m_{A'}>100$ MeV. For lower masses, there will be coherence across the nucleus.
This formula is rather approximate, and the hadronic matrix 
elements are estimated to be $f_q \sim 0.1$ \cite{Bramante:2016rdh}.

The mass splitting considered are always smaller than the mediator mass. The excited state can only decay through an off-shell $A'$ to $e^+ e^-$ if the splitting is above threshold, or through $3 \gamma$ which is a loop process. Both of these are highly suppressed. As a result some other decay mechanism might be required to deplete the excited state in the first place. We will assume that excited DM is long-lived on the detector scale. As a result, the observable is either ground state decay ($^{180m}$Ta) or $\gamma$ end-point ($^{178m}$Hf). For large statistics, we can also envision placing the isomer in the vicinity of a conventional DM detector. Thus if $\chi_1$ transitions into $\chi_2$ by scattering with the isomer, it now has enough energy to deposit its excess inside a conventional detector. 

To this end we plot in Fig.~\ref{fig2} the reach for 3 scattering events for various isomer target options as a function of the dark photon model parameters i.e. the mixing $\epsilon$ vs dark photon mass $m_{A'}$. Also shown are contours of the subsequent mean-free-path of the excited state $\chi_2$ in a typical DM detector with Xenon target. The DM mass $M_{\chi}=1$ TeV and the splitting is fixed to be $\delta M_{\chi}=500$ keV, such that existing inelastic DM limits from CRESST are absent. The elastic loop limits from Eqn.(\ref{elasticDF}) are shown in gray. Also shown in gray are limits on the dark photon mediator itself from Babar. 

The projections for $^{178m}$Hf and $^{180m}$Ta are not competitive due to small exposure and small energy threshold respectively. However $^{177m}$Lu as well as $^{137m}$Ba could probe new parameter space.

\subsection{Electric Dipole DM}
\begin{figure}[htpb]
\centering
\includegraphics[width=1\columnwidth,keepaspectratio]{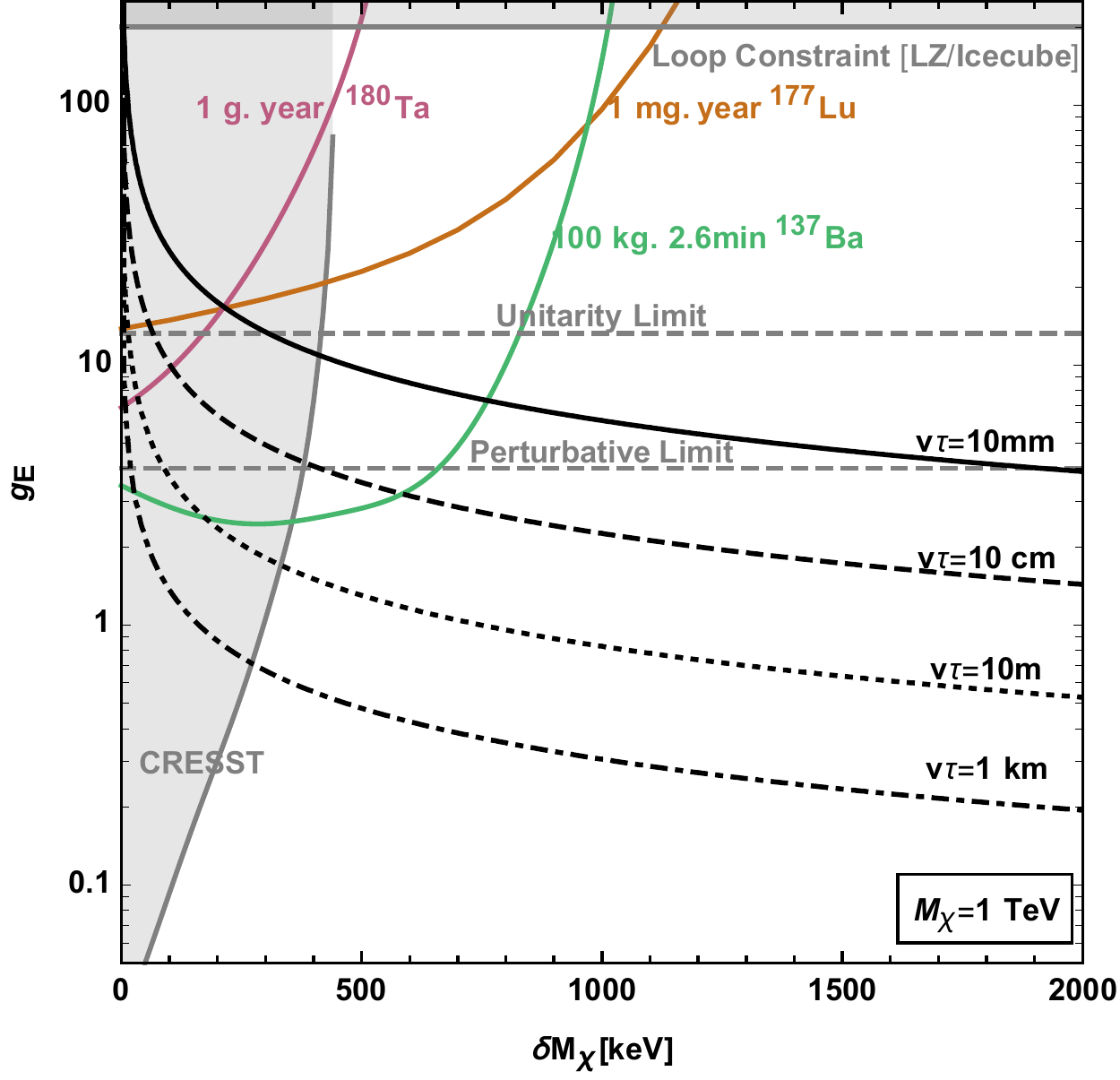}
\caption{Reach for Electric Dipole DM for different metastable isomer targets. Current limits on this model, from CRESST and the loop constraint from LZ/Icecube are also shown. The unitarity and perturbativity constraints are marked in dashed grey. Contours of constant lifetime of the excited DM state are also displayed. }
\label{fig3}
\end{figure}

While DM is unlikely to have an electric charge (unless it is very small), 
it may have a variety of electromagnetic form factors, including 
magnetic and electric dipoles \cite{Pospelov:2000bq}.
For electric dipole DM, we start with the Lagrangian\cite{Feldstein:2010su},
\begin{equation}
\mathcal{L} =i \frac{e g_E}{8M_{\chi}}\bar{\chi}_2\sigma^{\mu \nu}\chi_1 \tilde{F}^{\mu\nu} +{\rm h.c}.
\end{equation}
There are perturbativity and unitarity limits on $g_E$ derived in \cite{Sigurdson:2004zp}, and quoted here:
\begin{equation}
g_E <4 \left( \text{Perturbative}\right) \quad g_E < \frac{4}{e} \left( \text{Unitary}\right)
\end{equation}
The differential cross-section is \cite{Banks:2010eh},
\begin{equation}
\frac{d\sigma}{dq^2}=\frac{\pi \alpha^2}{v^2}\frac{g_E^2}{4 M_{\chi} ^2 } \frac{1}{q^2}.
\end{equation}
The elastic loop-limit can be estimated as \cite{Bramante:2016rdh}
\begin{equation}
\sigma_{\text{loop}}=\frac{\alpha^4_{\text{em}}}{\pi} \left(\frac{3 g_E^2}{16 M^2_\chi}\right)^2 \mu^2_n.
\end{equation}
Finally, the lifetime \cite{Feldstein:2010su} is given by
\begin{equation}
\Gamma(\chi_2 \rightarrow \chi_1 \gamma) = \alpha g_E^2 \frac{\delta M_{\chi}^3}{M^2_{\chi}}.
\end{equation}
In this model, the detection scheme involves the DM particle $\chi_1$ up-scattering off the metastable isomer to the $\chi_2$ state which decays within its characteristic decay length $v\tau$. Contours for 3 scatters are plotted in Fig.~\ref{fig3} for different nuclear targets. Also plotted are contours of constant $v\tau$ in order to illustrate the size of detector necessary to achieve detection of $\mathcal{O}(1)$ of the excited DM decays. We see that only $^{177}$Lu and $^{137}$Ba can set limits. Furthermore if perturbative/unitary constraints are taken into account, only $^{137}$Ba stays relevant. For the region of parameter space relevant, the decay is almost prompt, leading to a gamma ray corresponding to the mass-splitting. 

\section{Conclusions}
\label{conclusion}
Nuclear isomers appear to offer the unique opportunity to probe two generic classes of DM, namely, strongly interacting DM and inelastic DM. It has generally been difficult to constrain these models in direct detection experiments since the scattering process requires nuclei to impart energy to the DM in order to have an observable effect. By using the stored energy in the isomer, the methods proposed by us enable experimental access to this class of well motivated models. 

Our methods can potentially be implemented using existing resources. For example, by placing low threshold DM detectors near a suitable nuclear isomer source, one can search for either the re-scattering or decay of excited DM states subsequent to a scattering event with the isomer. The nuclear isomer sources can come from current investments in this area: either from naturally occurring tantalum, hafnium produced by the Department of Energy, barium produced from cesium in nuclear waste or lutetium from medical waste. Additionally, existing experimental setups that probe the lifetime of $^{180m}$Ta can also be used to look for the de-excitation caused by DM. 

While an unambiguous bound can be placed on the interaction of DM with specific nuclear isomers, there is theoretical uncertainty in translating these bounds into fundamental parameters in the Lagrangian since this requires knowledge of nuclear matrix elements. We have adopted an order of magnitude approach to this problem, encouraged by the fact that the DM scattering process transfers momenta $\sim 100$ MeV, which is at the same scale as the nuclei. This ought to enable transitions to occur without additional suppression. However, it would be preferable to directly measure these matrix elements, for example, through dedicated neutron scattering experiments.  This is likely to be of most use in the case of $^{180m}$Ta whose decay has never been observed preventing a comparison of our order of magnitude analysis with existing data. 

\appendix
\section{Nuclear States Data}
\label{append}
In this section we provide the $J$,$K$ quantum numbers as well as the $E_N$ used in calculating the scattering rates. All of this data was extracted from the Nudat2 database \cite{nudat20112}.

For tantalum, the isomeric state is in $\{9,9^+,77.2\}$ in $\{K,J^p,E_N~{\rm[keV]}\}$ format. The ground state is at $\{1,1^+,0\}$ and the only intermediate state is at $\{1,2^+,39.5\}$.

Caesium beta decays to an isomeric state of barium which is in $\{\frac{11}{2},\frac{11}{2}^-,661.7\}$. The ground state is at $\{\frac{1}{2},\frac{3}{2}^+,0\}$ and the only intermediate state is at $\{\frac{1}{2},\frac{1}{2}^+,283.5\}$.

There are several states between the isomeric and ground states in lutetium and hafnium, they are provided in Table.\ref{luttable} and Table.\ref{hftable} respectively.

\begin{table}[ht]
\begin{center}
 \begin{tabular}{||c|C{12mm}|c| |c|C{12mm}|c||} 
 \hline
$K$ & $J^p$ & $E_N$ [keV] & $K$ & $J^p$ & $E_N$ [keV]\\
\hline
\multirow{7}{10mm}{\centering $\frac{7}{2}$}&
%$\frac{7}{2}$
$\frac{7}{2}^{+}$&$0$& \multirow{5}{10mm}{\centering $\frac{1}{2}$}&

$\frac{1}{2}^{+}$
&$569.7$\\
%$\frac{7}{2}$
&$\frac{9}{2}^{+}$&$121.6$&
%$\frac{1}{2}$
&$\frac{3}{2}^{+}$&$573.6$
\\
%$\frac{7}{2}$
&$\frac{11}{2}^{+}$&$268.8$&
%$\frac{1}{2}$
&$\frac{5}{2}^{+}$&$709.5$
\\
%$\frac{7}{2}$
&$\frac{13}{2}^{+}$&$440.6$&
%$\frac{1}{2}$
&$\frac{7}{2}^{+}$&$720.8$
\\
%$\frac{7}{2}$
&$\frac{15}{2}^{+}$&$636.2$&
%$\frac{1}{2}$
&$\frac{9}{2}^{+}$&$956.6$
\\
\cline{4-6}
%$\frac{7}{2}$
&$\frac{17}{2}^{+}$&$854.3$&
\multirow{3}{10mm}{\centering$\frac{3}{2}$}&$\frac{3}{2}^{+}$&$760.8$
\\
\cline{1-3}
\multirow{5}{10mm}{\centering $\frac{9}{2}$}&$\frac{9}{2}^{-}$&$150.3$&
%$\frac{3}{2}$
&$\frac{5}{2}^{+}$&$822.0$
\\
%$\frac{9}{2}$
&$\frac{11}{2}^{-}$&$289.0$&
%$\frac{3}{2}$
&$\frac{7}{2}^{+}$&$906.7$
\\
\cline{4-6}
%$\frac{9}{2}$
&$\frac{13}{2}^{-}$&$451.5$&
\multirow{5}{10mm}{\centering$\frac{1}{2}$}&$\frac{5}{2}^{-}$&$761.7$
\\
%$\frac{9}{2}$
&$\frac{15}{2}^{-}$&$637.1$&
%$\frac{13}{2}$
&$\frac{1}{2}^{-}$&$795.2$
\\
%$\frac{9}{2}$
&$\frac{17}{2}^{-}$&$844.9$&
%$\frac{13}{2}$
&$\frac{9}{2}^{-}$&$811.5$
\\
\cline{1-3}
\multirow{5}{10mm}{\centering $\frac{5}{2}$}&$\frac{5}{2}^{+}$&$458.0$&
%$\frac{13}{2}$
&$\frac{3}{2}^{-}$&$956.5$
\\
%$\frac{5}{2}$
&$\frac{7}{2}^{+}$&$552.1$&
%\frac{13}{2}$
&$\frac{13}{2}^{-}$&$957.3$
\\
\cline{4-6}
%$\frac{5}{2}$
&$\frac{9}{2}^{+}$&$671.9$&
$\frac{23}{2}$&$\frac{23}{2}^{-}$&$970.2$
\\
%$\frac{5}{2}$
&$\frac{11}{2}^{+}$&$816.7$& & & 

\\
\hline
 \end{tabular}
 \end{center}
 
 \caption{$K$ and $J$ quantum numbers as well as $E_N$ the energy above the ground state are tabulated for $^{177}$Lu}
 \label{luttable}
 \end{table}

 \begin{table*}[ht]
 \begin{tabular}{c c c c}% c c}
 \begin{minipage}{.24\linewidth}
%\begin{center}
 \begin{tabular}{||c|C{12mm}|c|} 
 \hline
$K$ & $J^p$ & $E_N$ [keV] \\
\hline
\multirow{7}{10mm}{\centering $0$}&$0^{+}$&$0$\\
%$0$
&$2^{+}$&$93.2$\\ 
%$0$
&$4^{+}$&$306.6$\\
%$0$
&$6^{+}$&$632.2$\\ 
%$0$
&$8^{+}$&$1058.6$\\
%$0$
&$10^{+}$&$1570.3$\\
%$0$
&$12^{+}$&$2149.6$\\ 
\hline
\multirow{3}{*}{\centering$0$}&$0^{+}$&$1199.4$\\ 
%$0$
&$2^{+}$&$1276.7$\\
%0$
&$4^{+}$&$1450.4$\\ 
\hline
\multirow{4}{*}{\centering$0$}&$0^{+}$&$1434.2$\\ 
%$0$
&$2^{+}$&$1479.5$\\
%$0$
&$4^{+}$&$1636.6$\\ 
%$0$
&$6^{+}$&$1731.1$\\
\hline
\multirow{3}{*}{\centering$0$}&$0^{+}$&$1443.9$\\
%$0$
&$2^{+}$&$1513.6$\\ 
%$0$
&$4^{+}$&$1654.3$\\
\hline
\multirow{3}{10mm}{\centering$0$}
&$0^{+}$&$1772.2$\\
%$0$
&$2^{+}$&$1818.3$\\ 
%$0$
&$4^{+}$&$1956.4$\\

& & \\
\hline
\end{tabular}
\end{minipage}&
\begin{minipage}{.23\linewidth}
\begin{tabular}{|c|C{12mm}|c|} 
 \hline
$K$ & $J^p$ & $E_N$ [keV]\\
 \hline

\multirow{5}{10mm}{\centering$1$}&$1^{-}$&$1310.1$\\ 
%$1$
&$2^{-}$&$1362.6$\\
%$1$
&$3^{-}$&$1433.6$\\ 
%$1$
&$4^{-}$&$1538.8$\\
%$1$
&$5^{-}$&$1651.5$\\ 
\hline
\multirow{8}{*}{\centering$2$}&$2^{+}$&$1174.6$\\
%$2$
&$3^{+}$&$1268.5$\\
%$2$
&$4^{+}$&$1384.5$\\
%$2$
&$5^{+}$&$1533.2$\\
%$2$
&$6^{+}$&$1691.1$\\
%$2$
&$7^{+}$&$1890.0$\\ 
%$2$
&$8^{+}$&$2082.2$\\
%$2$
&$9^{+}$&$2315.8$\\ 
\hline
\multirow{5}{*}{\centering$2$}&$2^{-}$&$1260.2$\\ 
%$2$
&$3^{-}$&$1322.5$\\ 
%$2$
&$4^{-}$&$1409.4$\\ 
%$2$
&$5^{-}$&$1512.6$\\
%$2$
&$6^{-}$&$1648.8$\\
\hline
 $2$&$2^{+}$&$1561.5$\\
 & & \\
& & \\

 \hline
 \end{tabular}
\end{minipage}&
\begin{minipage}{.23\linewidth}
\begin{tabular}{|c|C{12mm}|c|} 
 \hline
$K$ & $J^p$ & $E_N$ [keV]\\
 \hline

\multirow{4}{10mm}{\centering$2$}&$2^{-}$&$1566.7$\\ 
%$2$
&$3^{-}$&$1639.7$\\
%$2$
&$4^{-}$&$1747.1$\\ 
%$2$
&$5^{-}$&$1863.7$\\
\hline
$2$&$2^{+}$&$1808.3$\\
\hline
\multirow{3}{*}{\centering$2$}&$2^{-}$&$1857.2$\\ 
%$2$
&$3^{-}$&$1917.4$\\
%$2$
&$4^{-}$&$2027.6$\\ 
\hline
\multirow{3}{*}{\centering$3$}
&$3^{+}$&$1758.0$\\
%$3$
&$4^{+}$&$1953.1$\\ 
%$3$
&$5^{+}$&$2068.1$\\
\hline
\multirow{2}{*}{\centering$3$}&$3^{-}$&$1803.4$\\ 
%$3$
&$4^{-}$&$1913.6$\\
\hline
\multirow{2}{*}{\centering$3$}&$3^{+}$&$1862.2$\\
%$3$
&$4^{+}$&$1869.8$\\ 
\hline
\multirow{6}{*}{\centering$4$}&$4^{+}$&$1513.8$\\ 
%$4$
&$5^{+}$&$1640.5$\\
%$4$
&$6^{+}$&$1788.6$\\ 
%$4$
&$7^{+}$&$1953.7$\\
%$4$
&$8^{+}$&$2154.1$\\ 
%$4$
&$9^{+}$&$2349.7$\\
\hline
\end{tabular}
\end{minipage}&
\begin{minipage}{.23\linewidth}
\begin{tabular}{|c|C{12mm}|c||} 
 \hline
$K$ & $J^p$ & $E_N$ [keV]\\
 \hline

\multirow{4}{10mm}{\centering$5$}&$5^{-}$&$1636.7$\\ 
%$5$
&$6^{-}$&$1781.3$\\
%$5$
&$7^{-}$&$1947.0$\\ 
%$5$
&$8^{-}$&$2137.4$\\
\hline
\multirow{5}{10mm}{\centering$6$}&$6^{+}$&$1554.0$\\ 
%$6$
&$7^{+}$&$1741.7$\\
%$6$
&$8^{+}$&$1952.0$\\ 
%$6$
&$9^{+}$&$2183.4$\\
%$6$
&$10^{+}$&$2433.7$\\ 
\hline
\multirow{6}{*}{\centering$8$}&$8^{-}$&$1147.4$\\
%$8$
&$9^{-}$&$1364.1$\\ 
%$8$
&$10^{-}$&$1601.5$\\ 
%$8$
&$11^{-}$&$1859.1$\\
%$8$
&$12^{-}$&$2136.5$\\
%$8$
&$13^{-}$&$2433.3$\\
\hline
\multirow{4}{10mm}{\centering
$8$}&$8^{-}$&$1479.0$\\
%$8$
&$9^{-}$&$1697.5$\\
 %$8$
 &$10^{-}$&$1939.1$\\
 %$8$
 &$11^{-}$&$2202.5$\\
 \hline
 $10$&$10^{+}$&$2440.2$\\
 \hline
$16$&$16^{+}$&$2446.1$\\
\hline
 \end{tabular}
 %\end{center}
 \end{minipage}
 \end{tabular}
 \caption{$K$ and $J$ quantum numbers as well as $E_N$ the energy above the ground state are tabulated for $^{178}$Hf}
 \label{hftable}
 \end{table*}
 \acknowledgements
 We would like to thank Bjoern Lehnert for useful explanations on the Tantalum decay experiment and Giovanni Benato for useful comments. M.P's Research at Perimeter
Institute is supported by the Government of Canada through Industry Canada and by the Province of Ontario through the Ministry of Economic Development and Innovation. S.R. was supported in part by the NSF under grants PHY- 1638509, the Simons Foundation Award 378243 and the Heising-Simons Foundation grants 2015-038 and 2018-0765. H.R. is supported in part by the DOE under contract DE-AC02-05CH11231. Some of this work was completed at the Aspen Center for Physics, which is supported by NSF grant PHY-1607611. This research was supported in part by the Munich Institute for Astro- and Particle Physics (MIAPP) of the DFG cluster of excellence "Origin and Structure of the Universe".
\newpage
\bibliographystyle{unsrt}
\bibliography{bibliography}

\begin{thebibliography}{10}

\bibitem{Hooper:2018bfw}
Dan Hooper and Samuel~D. McDermott.
\newblock {Robust Constraints and Novel Gamma-Ray Signatures of Dark Matter
  That Interacts Strongly With Nucleons}.
\newblock {\em Phys. Rev.}, D97(11):115006, 2018.

\bibitem{DeLuca:2018mzn}
Valerio De~Luca, Andrea Mitridate, Michele Redi, Juri Smirnov, and Alessandro
  Strumia.
\newblock {Colored Dark Matter}.
\newblock {\em Phys. Rev.}, D97(11):115024, 2018.

\bibitem{Neufeld:2018slx}
David~A. Neufeld, Glennys~R. Farrar, and Christopher~F. McKee.
\newblock {Dark Matter that Interacts with Baryons: Density Distribution within
  the Earth and New Constraints on the Interaction Cross-section}.
\newblock {\em Astrophys. J.}, 866(2):111, 2018.

\bibitem{TuckerSmith:2001hy}
David Tucker-Smith and Neal Weiner.
\newblock {Inelastic dark matter}.
\newblock {\em Phys. Rev.}, D64:043502, 2001.

\bibitem{Feldstein:2010su}
Brian Feldstein, Peter~W. Graham, and Surjeet Rajendran.
\newblock {Luminous Dark Matter}.
\newblock {\em Phys. Rev.}, D82:075019, 2010.

\bibitem{Pospelov:2013nea}
Maxim Pospelov, Neal Weiner, and Itay Yavin.
\newblock {Dark matter detection in two easy steps}.
\newblock {\em Phys. Rev.}, D89(5):055008, 2014.

\bibitem{Bramante:2016rdh}
Joseph Bramante, Patrick~J. Fox, Graham~D. Kribs, and Adam Martin.
\newblock {Inelastic frontier: Discovering dark matter at high recoil energy}.
\newblock {\em Phys. Rev.}, D94(11):115026, 2016.

\bibitem{Berestetsky:1982aq}
V.~B. Berestetskii, E.~M. Lifshitz, and L.~P. Pitaevskii.
\newblock {\em {QUANTUM ELECTRODYNAMICS}}, volume~4 of {\em Course of
  Theoretical Physics}.
\newblock Pergamon Press, Oxford, 1982.

\bibitem{karamian1999fast}
SA~Karamian, CB~Collins, JJ~Carroll, J~Adam, AG~Belov, and VI~Stegailov.
\newblock Fast neutron induced depopulation of the 180 ta m isomer.
\newblock {\em Physical Review C}, 59(2):755, 1999.

\bibitem{hartouni2008theoretical}
EP~Hartouni, M~Chen, MA~Descalle, JE~Escher, A~Loshak, P~Navratil, WE~Ormand,
  J~Pruet, IJ~Thompson, and TF~Wang.
\newblock Theoretical assessment of 178m2hf de-excitation.
\newblock Technical report, Lawrence Livermore National Lab.(LLNL), Livermore,
  CA (United States), 2008.

\bibitem{karamian2011cross}
SA~Karamian and JJ~Carroll.
\newblock Cross section for inelastic neutron``acceleration''by 178 hf m 2.
\newblock {\em Physical Review C}, 83(2):024604, 2011.

\bibitem{Lehnert:2016iku}
Björn Lehnert, Mikael Hult, Guillaume Lutter, and Kai Zuber.
\newblock {Search for the decay of nature's rarest isotope $^{\rm {180m}}$Ta}.
\newblock {\em Phys. Rev.}, C95(4):044306, 2017.

\bibitem{forthcoming}
B.~Lehnert, M.~Hult, G.~Lutter, M.Pospelov, S.Rajendran, H.Ramani, and K.~Zube.
\newblock First limits on dark matter accelerated by $^{180m}$tantalum.
\newblock {\em forthcoming}, 2019.

\bibitem{Ejiri:2017dro}
H.~Ejiri and T.~Shima.
\newblock {K-hindered beta and gamma transition rates in deformed nuclei and
  the halflife of (180)Ta( )m( )}.
\newblock {\em J. Phys.}, G44(6):065101, 2017.

\bibitem{Engel:1999kv}
J.~Engel and P.~Vogel.
\newblock {Neutralino inelastic scattering with subsequent detection of nuclear
  gamma-rays}.
\newblock {\em Phys. Rev.}, D61:063503, 2000.

\bibitem{sandor1993shape}
RKJ Sandor, HP~Blok, M~Girod, MN~Harakeh, CW~de~Jager, V~Yu Ponomarev, and
  H~de~Vries.
\newblock Shape transition of 146nd deduced from an inelastic
  electron-scattering experiment.
\newblock {\em Nuclear Physics A}, 551(3):378--408, 1993.

\bibitem{nudat20112}
NuDat NuDat.
\newblock 2.5 database http://www. nndc. bnl. gov/nudat2/, in: National nuclear
  data center.
\newblock {\em Brookhaven National Laboratory}, 2011.

\bibitem{Landau:1991wop}
Lev~Davidovich Landau and E.~M. Lifshits.
\newblock {\em {Quantum Mechanics}}, volume v.3 of {\em Course of Theoretical
  Physics}.
\newblock Butterworth-Heinemann, Oxford, 1991.

\bibitem{landau1981course}
LD~Landau, EM~Lifshitz, and LP~Pitaevskij.
\newblock {\em Course of theoretical physics. vol. 10: Physical kinetics}.
\newblock Oxford, 1981.

\bibitem{Hall:1997ah}
Lawrence~J. Hall, Takeo Moroi, and Hitoshi Murayama.
\newblock {Sneutrino cold dark matter with lepton number violation}.
\newblock {\em Phys. Lett.}, B424:305--312, 1998.

\bibitem{Izaguirre:2017bqb}
Eder Izaguirre, Yonatan Kahn, Gordan Krnjaic, and Matthew Moschella.
\newblock {Testing Light Dark Matter Coannihilation With Fixed-Target
  Experiments}.
\newblock {\em Phys. Rev.}, D96(5):055007, 2017.

\bibitem{Batell:2009vb}
Brian Batell, Maxim Pospelov, and Adam Ritz.
\newblock {Direct Detection of Multi-component Secluded WIMPs}.
\newblock {\em Phys. Rev.}, D79:115019, 2009.

\bibitem{Pospelov:2000bq}
Maxim Pospelov and Tonnis ter Veldhuis.
\newblock {Direct and indirect limits on the electromagnetic form-factors of
  WIMPs}.
\newblock {\em Phys. Lett.}, B480:181--186, 2000.

\bibitem{Sigurdson:2004zp}
Kris Sigurdson, Michael Doran, Andriy Kurylov, Robert~R. Caldwell, and Marc
  Kamionkowski.
\newblock {Dark-matter electric and magnetic dipole moments}.
\newblock {\em Phys. Rev.}, D70:083501, 2004.
\newblock [Erratum: Phys. Rev.D73,089903(2006)].

\bibitem{Banks:2010eh}
Tom Banks, Jean-Francois Fortin, and Scott Thomas.
\newblock {Direct Detection of Dark Matter Electromagnetic Dipole Moments}.
\newblock 2010.

\end{thebibliography}

\end{document}